\documentclass[journal]{IEEEtran}
\usepackage{float}
\usepackage{stfloats}
\usepackage{graphicx}
\usepackage{makeidx}
\usepackage{amsmath}
\usepackage{makecell}
\usepackage{amsfonts}
\usepackage{bbding}
\usepackage{amssymb}
\usepackage{wrapfig}
\usepackage{psfrag}
\usepackage{indentfirst}
\usepackage{cite}
\usepackage[font=small,labelsep=period]{caption}
\usepackage{bm}
\usepackage[ruled]{algorithm2e}
\usepackage{algorithmic}
\usepackage{threeparttable}
\usepackage{color}
\usepackage{subfig}
\usepackage{array,color}
\usepackage{amsfonts,amssymb}
\usepackage{CJK}
\usepackage{tikz}
\usepackage{etoolbox}
\usepackage[figuresright]{rotating}
\usepackage{epsfig}
\usepackage{epstopdf}
\usepackage{mdwlist}
\usepackage{multirow}
{\begin{list}{}{\settowidth{\labelwidth}{\bf #1}%
			\setlength{\leftmargin}{\labelwidth}%
			\addtolength{\leftmargin}{\labelsep}%
			}}%
	{\end{list}}

\newcommand{\tabincell}[2]{\begin{tabular}{@{}#1@{}}#2\end{tabular}} 

\ifCLASSINFOpdf
\else
\fi

\DeclareRobustCommand*{\IEEEauthorrefmark}[1]{%
	\raisebox{0pt}[0pt][0pt]{\textsuperscript{\footnotesize\ensuremath{#1}}}}

\begin{document}

\title{Symbol-Based Multi-Layer Iterative Successive Interference Cancellation for Faster-Than-Nyquist Signaling}


\author{\IEEEauthorblockN{Qiang Li\IEEEauthorrefmark{1}\IEEEauthorrefmark{2},
		Feng-Kui Gong\IEEEauthorrefmark{1}\IEEEauthorrefmark{2}, Pei-Yang Song\IEEEauthorrefmark{1} and Sheng-Hua Zhai\IEEEauthorrefmark{3}\IEEEauthorrefmark{4}}
	
	\IEEEauthorblockA{
		\IEEEauthorrefmark{1} State Key Laboratory of Integrated Service Networks (ISN), Xidian University, Shaanxi, Xi'an 710071, China \\
		\IEEEauthorrefmark{2} Collaborative innovation center of information sensing  and understanding, Shaanxi, Xi'an 710071, China \\
		\IEEEauthorrefmark{3} School of Information and Electronics, Beijing Institute of Technology, Beijing 100081, China \\
		\IEEEauthorrefmark{4} CAST-Xi'an Institute of Space Radio Technology, Xi'an 710071, China \\		
					}
		}

\maketitle

\begin{abstract}
	
A symbol-based multi-layer iterative successive interference cancellation (MLISIC) algorithm is proposed to eliminate the inter-symbol interference (ISI) for faster-than-Nyquist (FTN) signaling. The computational complexity of the proposed MLISIC algorithm is much lower than most existing block-based estimation algorithms. By comparison with existing symbol-based algorithms, the proposed MLISIC algorithm has higher estimation accuracy. Furthermore, by means of utilizing more accurate symbols and elaborate length of successive interference cancellation, an improved MLISIC (IMLISIC) algorithm is presented to further promote the estimation accuracy. Simulation results show in mild ISI cases, the proposed MLISIC and IMLISIC algorithms can approximate the theoretical performance even using 256-amplitude phase shift keying (APSK) which is adopted in digital video broadcasting-satellite-second generation extension (DVB-S2X), i.e., bit error rate (BER) performance degradation is about 0.03 dB when the signal-to-noise ratio is high. In other words, using the proposed MLISIC and IMLISIC algorithms in mild ISI cases, the spectrum efficiency can be improved with negligible BER performance degradation. In moderate ISI cases, BER performance degradation of our proposed IMLISIC algorithm is no more than 0.13 dB when adopting 64/128/256-APSK, which is still satisfying. Besides, compared with existing symbol-based algorithms, the greater BER performance improvement can be achieved under severer ISI circumstances. 

\end{abstract}


%
\IEEEpeerreviewmaketitle

\section{Introduction}
%
\IEEEPARstart{W}{ith} the rapid development of the wireless communication technology in recent years, wireless devices grow exponentially, which results in more and more demand for communication services. In addition, the high-throughput satellite becomes a trend due to the scarcity of spectrum resources in satellite communications. High-order modulations such as 256-amplitude phase shift keying (APSK) and low rolling factors such as 0.05 are recommended in digital video broadcasting-satellite-second generation extension (DVB-S2X) \cite{DVB-S2X}, however, they are sensitive to the phase noise distortion and timing error respectively. As a non-orthogonal transmission scheme, faster-than-Nyquist (FTN) signaling has attracted much more attention.

When designing the traditional communication system, the Nyquist criterion must be followed to avoid the inter-symbol interference (ISI). Nevertheless, the orthogonality of Nyquist signaling is at the cost of sacrificing the spectrum efficiency. By deliberately introducing ISI, higher transmission rate and spectrum efficiency can be supported in the FTN system. Besides, FTN transmission can moderate the need for high-order modulations and low rolling factors. Thus, the adverse effect of the phase noise distortion and the timing error can be mitigated.  

FTN signaling was first proposed by Mazo in 1970s. It is demonstrated in \cite{6772210} that as long as the time acceleration parameter $\tau$ is no less than 0.802, i.e., $0.802 \le \tau  \le 1$, the minimum distance of uncoded binary sinc pulse transmission is not reduced in FTN signaling. Compared with the traditional Nyquist system, the FTN system can transmit about 25\% more symbols in the same bandwidth without bit error rate (BER) performance degradation. The calculation of the minimum distance for FTN signaling was then given in \cite{21281}. Due to the fact that the pulse response of sinc pulse is infinite in time domain, raised-cosine (RC) pulses are introduced for FTN signaling in \cite{1231648}. Since then, FTN signaling has been widely investigated in different aspects \cite{7845669}, such as FTN signaling using non-binary or high-order modulations \cite{4524864,8408541}. The constrained capacity for FTN signaling is also clarified in \cite{4777625}, which shows that the FTN system can achieve a higher capacity than the Nyquist system unless sinc pulse is adopted. The advantages of frequency-domain FTN systems can be found in \cite{1523482,8276223}. In addition, FTN signaling has been applied to more scenarios, such as multi-carrier \cite{5645721,4939227,8440066} and multi-input-multi-output (MIMO) \cite{4801456,8405769} systems. Especially for satellite communications, FTN signaling is investigated as an innovative method to improve spectrum efficiency in \cite{5288497,6559982}. A receiver architecture for FTN signaling in the digital video broadcasting-satellite-second generation (DVB-S2) standard \cite{DVB-S2} is also presented \cite{7391110}.

On account of the violation of the Nyquist criterion in FTN system, it is inevitable to introduce undesired ISI. Hence, lots of research literature focused on the ISI cancellation algorithms. As known, the optimal ISI cancellation algorithm is the maximum likelihood sequence estimation (MLSE), which is not practical due to its non-deterministic polynomial (NP)-hard computational complexity. Based on the M-algorithm and the Bahl-Cocke-Jelinek-Raviv (BCJR) algorithm, J. B. Anderson proposed the M-BCJR algorithm for FTN signaling in \cite{6241379}. Since the M-BCJR algorithm only keeps several maximum states as possible states at each time, it can maintain the good performance when reducing the computational complexity of the BCJR algorithm. But the computational complexity of the M-BCJR algorithm is still prohibitive for the practical implementation. In \cite{1523482,4595029}, the truncated Viterbi algorithm and M-algorithm are also investigated for the uncoded single-carrier FTN system. In 2013, a low complexity and effective frequency-domain equalizer (FDE) for the FTN system is proposed \cite{6574905}. To a certain extent, however, its spectrum efficiency is decreased because of the insertion of guard interval, e.g., cyclic prefix. As a commendable method to eliminate ISI, the performance of precoding in the FTN system can be found in \cite{7284519,7480840,8401480}. An iterative FTN detector in the presence of phase noise and carrier frequency offset is also developed in a factor graph framework \cite{201725}. Two successive symbol-by-symbol sequence estimators are proposed to estimate binary and 4-quadrature amplitude modulation (QAM) FTN signaling in \cite{7886296}. It is proved that one of these two sequence estimators, which is named after successive symbol-by-symbol with go-back-$\emph{K}$ sequence estimator (SSSgb$\emph{K}$SE), performs well in mild ISI cases and can significantly increase the transmission rate and spectrum efficiency. The semi-definite relaxation-based sequence estimation presented in \cite{7990502} shows pretty good performance, but its computational complexity is disappointing when the block length is not that small.

Inspired by the successive symbol-by-symbol sequence estimators in \cite{7886296}, we present a multi-layer iterative successive interference cancellation (MLISIC) algorithm in this paper. And then, the improved MLISIC (IMLISIC) algorithm is proposed to further promote the estimation accuracy. The main contributions of this paper are summarized as follows:
\begin{enumerate}
  \item Firstly, in consideration of the promising prospect of FTN signaling in satellite communications, FTN signaling is extended to high-order APSK modulations, which are adopted in DVB-S2X, in this paper. As far as we know, this is the first effort in the literature reported. In addition, all the modulation types adopted in DVB-S2X are evaluated in FTN systems.
  \item Secondly, we propose a new MLISIC algorithm. The proposed MLISIC algorithm is based on the reception of one single symbol, instead of a block. Hence, in contrast to most of the block-based sequence estimation algorithms mentioned above, lower processing delay can be achieved. Since the MLISIC algorithm is realized in time domain, it is not necessary to insert guard interval in FTN signaling, which ensures the high spectrum efficiency of FTN signaling. Besides, the computational complexity of our proposed MLISIC algorithm is quite lower than those of most block-based sequence estimation algorithms. Even compared with the SSSgb$\emph{K}$SE algorithm in \cite{7886296}, the proposed MLISIC algorithm has lower computational complexity in most cases. 
  \item Thirdly, based on the MLISIC algorithm, the IMLISIC algorithm is also proposed by using more accurate symbols and elaborate length of successive interference cancellation. The IMLISIC algorithm has higher estimation accuracy than the MLISIC algorithm. Hence, the proposed IMLISIC algorithm can achieve better BER performance than the proposed MLISIC algorithm even when their computational complexities are equal. 
  \item Fourthly but not the last, simulation results show that our proposed algorithms are characteristic by higher estimation accuracy than the FDE algorithm in \cite{6574905} and the SSSgb$\emph{K}$SE algorithm in \cite{7886296} in mild and moderate ISI cases. For all the modulation types adopted in DVB-S2 and DVB-S2X, our proposed algorithms can approximate the theoretical performance in mild ISI cases. Even under moderate ISI circumstances, for which the SSSgb$\emph{K}$SE algorithm is not suitable, the proposed algorithms perform quite well.
\end{enumerate}


The rest of this paper is organized as follows: Section \uppercase\expandafter{\romannumeral2} introduces the system model of FTN signaling. The constrained capacity for FTN signaling is also covered in Section \uppercase\expandafter{\romannumeral2}. Section \uppercase\expandafter{\romannumeral3} discusses the proposed MLISIC and IMLISIC algorithms in detail. Simulation results are demonstrated in Section \uppercase\expandafter{\romannumeral4}. Section \uppercase\expandafter{\romannumeral5} makes a brief conclusion.

\section{FTN SYSTEM MODEL AND The Constrained Capacity }

In order to better illustrate our proposed algorithms, the FTN system model is first introduced in this section. After that, the relation among Shannon capacity, the capacity of FTN system and the capacity of Nyquist system is briefly explained.

\subsection{FTN System Model}

\begin{figure}[!h]
\centering
\includegraphics[width=8.3cm]{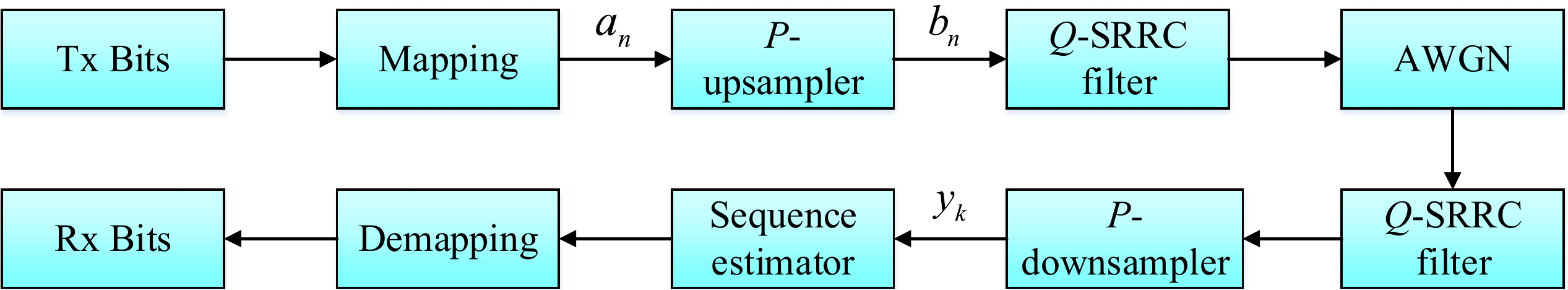}
\caption{The FTN system model}
\label{systemModel}
\end{figure}

The FTN system model used in this paper is illustrated in Fig. \ref{systemModel}, where AWGN indicates the additive white Gaussian noise. The FTN signaling shaping is composed of a $\emph{P}-$upsampler and a $\emph{Q}-$square root raised cosine (SRRC) shaping filter. $\emph{P}$ and $\emph{Q}$ represent the times of upsampling or downsampling used for the corresponding modules. As described in \cite{6692689}, the corresponding time acceleration parameter can be formulated as $\tau {\rm{ = }}{P \mathord{\left/ {\vphantom {P Q}} \right. \kern-\nulldelimiterspace} Q},P \le Q$, where ${ \cdot  \mathord{\left/ {\vphantom { \cdot   \cdot }} \right. \kern-\nulldelimiterspace}  \cdot }$ indicates the division operation. It is worth mentioning that if $\emph{P}$ equals to $\emph{Q}$, the system is the traditional Nyquist system.

Symbols after the $\emph{P}-$upsampler can be expressed as
\begin{equation}
{b_n} =
\left\{
\begin{array}{ll}
{{a_{{n \mathord{\left/ {\vphantom {n P}} \right. \kern-\nulldelimiterspace} P}}}} & {n = 0,P, \ldots ,\left( {N - 1} \right)P} \\
0 & {{\rm{otherwise}}}
\end{array},
\right.
\end{equation}
where $\emph{N}$ represents the total number of transmitted symbols. ${a_n},0 \le n \le N-1$ is the independent and identically distributed data symbols.

The transmitted signal $s\left( t \right)$ of the FTN system shown in Fig. \ref{systemModel} can be written as
\begin{equation}\label{eq:0}
\begin{split}
s\left( t \right) &= \sqrt {\tau {E_s}} \sum\nolimits_{n = 0}^{PN - 1} {{b_n}p\left( {t - n\tau T} \right)} \\
&= \sqrt {\tau {E_s}} \sum\nolimits_{n = 0}^{N - 1} {{a_n}p\left( {t - n\tau {T_s}} \right)} ,\;0 < \tau  \le 1,
\end{split}
\end{equation}
where $E_s$ is the average energy of the transmitted symbols. $p\left( t \right)$ is a unit energy pulse. ${1 \mathord{\left/ {\vphantom {1 T}} \right. \kern-\nulldelimiterspace} T}$ is the sampling rate after the upsampler. ${1 \mathord{\left/ {\vphantom {1 T_s}} \right. \kern-\nulldelimiterspace} T_s}$ and ${1 \mathord{\left/{\vphantom {1 {\tau T_s}}} \right.\kern-\nulldelimiterspace} {\tau T_s}}$ indicate the symbol rates of the Nyquist and FTN systems, respectively.

The received signal after the SRRC matching filter can be written as
\begin{flalign}
y\left( t \right) = \sqrt {\tau {E_s}} \sum\nolimits_{n = 0}^{N - 1} {{a_n}g\left( {t - n\tau {T_s}} \right) + w\left( t \right)} ,
\end{flalign}
where $g\left( t \right)=\int {p\left( x \right)p\left( {x - t} \right)dx},$ $w\left( t \right) = \int {n\left( x \right)p\left( {x - t} \right)} dx$. $n\left( t \right)$ is the zero-mean AWGN with variance ${\sigma ^2}$.

Assuming perfect timing synchronization between the transmitter and the receiver, symbols after the downsampler can be denoted as \cite{7990502}
\begin{equation}\label{eq:2}
\begin{split}
{y_k} &= y\left( {k\tau T_s} \right)  \\
&  = \sqrt {\tau {E_s}} \sum\nolimits_{n = 0}^{N-1} {{a_n}g\left( {k\tau T_s - n\tau T_s} \right) + w\left( {k\tau T_s} \right)}  \\
&  = \underbrace {\sqrt {\tau {E_s}} {a_k}g\left( 0 \right)}_{{\rm{desired}}\;{\rm{symbol}}} + \underbrace {\sqrt {\tau {E_s}} \sum\nolimits_{n = 0,n \ne k}^{N-1} {{a_n}g\left( {\left( {k - n} \right)\tau {T_s}} \right)} }_{{\rm{ISI}}}\\
&\quad + w\left( {k\tau T_s} \right), \\
\end{split}
\end{equation}
where $\sqrt {\tau {E_s}} {a_k}g\left( 0 \right)$ is the desired symbol. The latter item represents the ISI caused by adjacent symbols.

\begin{figure}[!h]
	\centering
	\includegraphics[width=8.8cm]{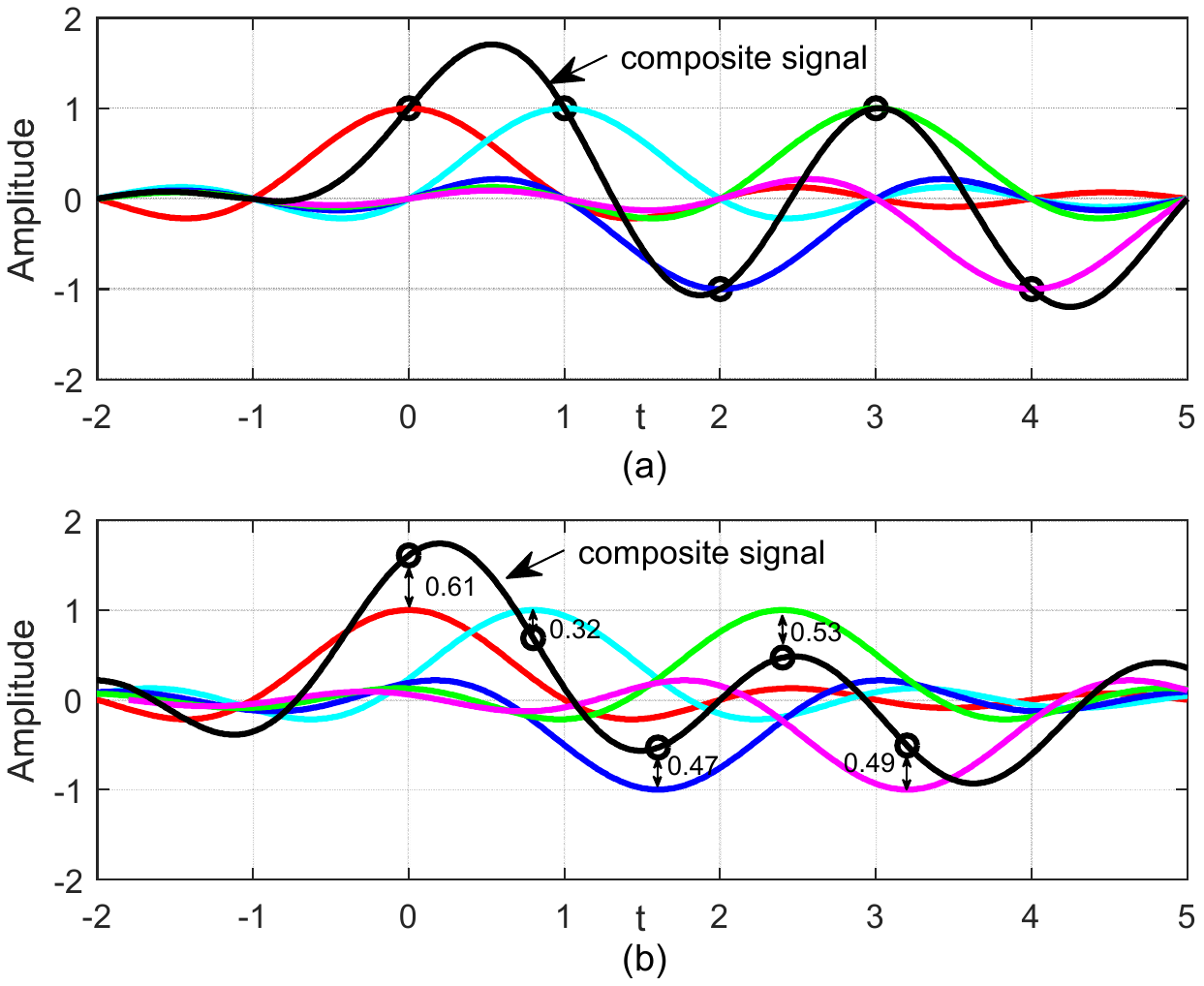}
	\caption{Receiving symbols in the noiseless (a) Nyquist system ($\tau  = 1$) and (b) FTN system ($\tau  = 0.8$), which use the same sinc pulse.}
	\label{ISI}
\end{figure}

When transmitted symbols are $\left\{ {1,1, - 1,1, - 1} \right\}$ and sinc pulse is adopted, the received symbols in the noiseless Nyquist and FTN ($\tau  = 0.8$) systems are shown in Fig. \ref{ISI}, where numbers such as 0.61 and 0.32 represent values of ISI.

\subsection{Constrained Capacity of the FTN System}

As described in \cite{6773024}, Shannon defined the capacity of the AWGN channel as
\begin{flalign}
\label{ShannonCapacity}
C = \int\limits_0^W {{{\log }_2}\left[ {1 + \frac{{2\bar P}}{{{N_0}}}{{\left| {P\left( f \right)} \right|}^2}} \right]} df,
\end{flalign}
where $W$ and ${\bar P}$ represent the one-sided bandwidth and the average power of transmitted signals, respectively. $N_0$ is the power spectrum density (PSD) of AWGN. $P\left( f \right)$ is the Fourier transform of the shaping pulse $p\left( t \right)$.

According to \cite{4777625}, the capacity of FTN system under the constraint ${W_{{\rm{FTN}}}} = {1 \mathord{\left/{\vphantom {1 {2\tau {T_s}}}} \right. \kern-\nulldelimiterspace} {2\tau {T_s}}}$ can be denoted as
\begin{flalign}
{C_{{\rm{FTN}}}} = \int\limits_0^{{W_{{\rm{FTN}}}}} {{{\log }_2}\left[ {1 + \frac{{2\bar P}}{{{N_0}}}{{\left| {P\left( f \right)} \right|}^2}} \right]} df.
\end{flalign}

Correspondingly, the capacity of Nyquist system is
\begin{flalign}
\label{NyquistCapacity}
{C_{\rm{N}}} = \int\limits_0^{{1 \mathord{\left/{\vphantom {1 {2{T_s}}}} \right.
\kern-\nulldelimiterspace} {2{T_s}}}} {{{\log }_2}\left[ {1 + \frac{{2\bar P{T_s}}}{{{N_0}}}} \right]} df.
\end{flalign}

Hence, it can be concluded that the constrained capacity of FTN system equals to the AWGN capacity defined by Shannon. Besides, the capacity of FTN system is no less than that of the corresponding Nyquist system, i.e., ${C_{{\rm{FTN}}}} \ge {C_{\rm{N}}}$. And ${C_{{\rm{FTN}}}} = {C_{\rm{N}}}$ only when $p\left( t \right)$ is the sinc pulse.

\begin{figure*}[!b]
\begin{align*} \label{eq3:MLISIC}
\begin{split}
{{\hat a}_{K',k - K'\left( {L - 1} \right)}} &= {\rm{deci}}\left\{ {{y_{k - K'\left( {L - 1} \right)}}} \right. \\
&- \left( {{G_{1,L}}{{\hat a}_{K' - 1,k - \left( {K' + 1} \right)\left( {L - 1} \right)}} +  \ldots  + {G_{1,2}}{{\hat a}_{K' - 1,k - K'\left( {L - 1} \right) - 1}}} \right) \\
&\left. { - \left( {{G_{1,2}}{{\hat a}_{K' - 1,k - K'\left( {L - 1} \right) + 1}} +  \ldots  + {G_{1,L}}{{\hat a}_{K' - 1,k - \left( {K' - 1} \right)\left( {L - 1} \right)}}} \right)} \right\}. \\
\end{split}
\tag{14}
\end{align*}
\end{figure*}

\section{Proposed Multi-layer Iterative Successive Interference Cancellation}

Given that our proposed algorithms are inspired by the successive symbol-by-symbol sequence estimators in \cite{7886296}, the SSSgb$\emph{K}$SE algorithm is first introduced in this section. And then, our proposed algorithms are explicated in detail. The comparison of computational complexity is presented at the end of this section.

\subsection{SSSgbKSE in \cite{7886296}}

Based on the FTN system model in Section \uppercase\expandafter{\romannumeral2}, symbols after the downsampler can be re-written as
\begin{equation}
\label{eq:G}
\begin{split}
{y_k} &= \underbrace {{G_{1,L}}{a_{k - L + 1}} +  \ldots  + {G_{1,2}}{a_{k - 1}}}_{{\rm{ISI}}\;{\rm{caused}}\;{\rm{by}}\;{\rm{previous}}\;{\rm{symbols}}}  \\
& + \underbrace {{G_{1,1}}{a_k}}_{{\rm{desired}}\;{\rm{symbol}}}  \\
& + \underbrace {{G_{1,2}}{a_{k + 1}} +  \ldots  + {G_{1,L}}{a_{k + L - 1}}}_{{\rm{ISI}}\;{\rm{caused}}\;{\rm{by}}\;{\rm{latter}}\;{\rm{symbols}}}, \\
\end{split}
\end{equation}
where ${G_{n,n'}} = g\left( {\left( {n - n'} \right)\tau {T_s}} \right)$.

Therefore, in order to estimate the symbol ${a_k}$, the interference of $L - 1$ symbols before the receiving symbol ${y_k}$ should be eliminated. Based on this idea, the estimation of the successive symbol-by-symbol sequence estimator (SSSSE) in \cite{7886296} can be expressed as
\begin{flalign}
{\hat a_k} = {\rm{deci}}\left\{ {{y_k} - \left( {{G_{1,L}}{{\hat a}_{k - L + 1}} +  \ldots  + {G_{1,2}}{{\hat a}_{k - 1}}} \right)} \right\},
\end{flalign}
where ${\mathop{\rm deci}\nolimits} \left( {x} \right)$ rounds $x$ to the nearest symbol in the constellation.

The computational complexity of the SSSSE algorithm is quite low, however, it is sensitive to the error propagation, which means the estimation accuracy of the latter symbols will be seriously affected by the incorrectly estimated symbol. In \cite{7886296}, the SSSgb$\emph{K}$SE algorithm is presented to reduce the error propagation effect and improve the estimation accuracy. The first step of the SSSgb$\emph{K}$SE algorithm is consistent with the SSSSE algorithm. Once ${\hat a_k}$ is estimated for the first time, the previous $\emph{K}$ estimated symbols, i.e., $\left\{ {{{\hat a}_{k - 1}}, \ldots ,{{\hat a}_{k - K}}} \right\},K \le L - 1$, can be re-estimated by using the current estimated symbol ${\hat a_k}$. With the re-estimated symbols $\left\{ {{{\tilde a}_{k - 1}}, \ldots ,{{\tilde a}_{k - K}}} \right\},K \le L - 1$, ${\hat a_k}$ can be re-estimated as
\begin{equation}
\label{eq:R1}
\begin{split}
{{\tilde a}_k} &= {\rm{deci}}\left\{ {{y_k} - \left( {\underbrace {{G_{1,L}}{{\hat a}_{k - L + 1}} +  \ldots }_{L - 1 - K\;{\rm{symbols}}\;{\rm{that}}\;{\rm{are}}\;{\rm{not}}\;{\rm{re - estimated}}}} \right.} \right. \\
&\left. {\left. { + \underbrace {{G_{1,K + 1}}{{\tilde a}_{k - K}} +  \ldots  + {G_{1,2}}{{\tilde a}_{k - 1}}}_{{\rm{previous}}\;K\;{\rm{re - estimated}}\;{\rm{symbols}}}} \right)} \right\}, \\
\end{split}
\end{equation}
where ${\tilde a_k}$ represents the re-estimated symbol. It can be concluded that for the SSSgb$\emph{K}$SE algorithm, each symbol is estimated up to $K + 2$ times.

\begin{figure*}[!b]
	\begin{align*} \label{eq3:IMLISIC}
	\begin{split}
	{{\hat a}_{K',k - \sum\nolimits_{i = 1}^{K'} {\left( {{L_i} - 1} \right)} }} &= {\rm{deci}}\left\{ {{y_{k - \sum\nolimits_{i = 1}^{K'} {\left( {{L_i} - 1} \right)} }}} \right. \\
	&- \left( {{G_{1,{L_{K'}}}}{{\hat a}_{K',k - \sum\nolimits_{i = 1}^{K'} {\left( {{L_i} - 1} \right)}  - \left( {{L_{K'}} - 1} \right)}} +  \ldots  + {G_{1,2}}{{\hat a}_{K',k - \sum\nolimits_{i = 1}^{K'} {\left( {{L_i} - 1} \right)}  - 1}}} \right) \\
	&\left. { - \left( {{G_{1,2}}{{\hat a}_{K' - 1,k - \sum\nolimits_{i = 1}^{K'} {\left( {{L_i} - 1} \right)}  + 1}} +  \ldots  + {G_{1,{L_{K'}}}}{{\hat a}_{K' - 1,k - \sum\nolimits_{i = 1}^{K'} {\left( {{L_i} - 1} \right)}  + \left( {{L_{K'}} - 1} \right)}}} \right)} \right\}. \\
	\end{split}
	\tag{17}
	\end{align*}
\end{figure*}

The formula used for re-estimating ${\hat a_{k - K'}},1 \le K' \le K$ can be expressed as
\begin{equation}
\label{eq:R2}
\begin{split}
{{\tilde a}_{k - K'}} &= \\
&{\rm{deci}}\left\{ {{y_{k - K'}} - \left( {\underbrace {{G_{1,L}}{{\hat a}_{k - K' - L + 1}} +  \ldots }_{L - 1\;{\rm{symbols}}\;{\rm{that}}\;{\rm{are}}\;{\rm{not}}\;{\rm{re - estimated}}}} \right)} \right. \\
&- \left. {\left( {\underbrace {{G_{1,2}}{{\tilde a}_{k - K' + 1}} +  \ldots  + {G_{1,K'}}{{\tilde a}_{k - 1}}}_{{\rm{previous}}\;K' - 1\;{\rm{re - estimated}}\;{\rm{symbols}}} + {G_{1,K' + 1}}{{\hat a}_k}} \right)} \right\}. \\
\end{split}
\end{equation}

\subsection{Proposed MLISIC}

As depicted in (\ref{eq:G}), received symbols can be estimated correctly as long as the ISI caused by adjacent symbols is completely eliminated. Therefore, once ${y_k}$ is received, the desired symbol ${a_{k - L + 1}}$ can be estimated for the first time as
\begin{equation}
\label{eq1:MLISIC}
\begin{split}
{{\hat a}_{1,k - L + 1}} &= \\
&{\rm{deci}}\left\{ {{y_{k - L + 1}} - \underbrace {\left( {{G_{1,L}}{y_{k - 2L + 2}} +  \ldots  + {G_{1,2}}{y_{k - L}}} \right)}_{{\rm{ISI}}\;{\rm{caused}}\;{\rm{by}}\;{\rm{previous}}\;L - 1\;{\rm{symbols}}}} \right. \\
&\left. { - \underbrace {\left( {{G_{1,2}}{y_{k - L + 2}} +  \ldots  + {G_{1,L}}{y_k}} \right)}_{{\rm{ISI}}\;{\rm{caused}}\;{\rm{by}}\;{\rm{latter}}\;L - 1\;{\rm{symbols}}}} \right\},\\
\end{split}
\end{equation}
which is the first step of the proposed MLISIC algorithm. It is worth mentioning that all the symbols used on the right side of (\ref{eq1:MLISIC}) are received symbols.

Once ${\hat a_{1,k - L + 1}}$ is obtained, the symbols after the first estimation ${\hat a_{1,k'}},k - 3L + 3 \le k' \le k - L + 1,k' \ne k - 2L + 2$ can be used to re-estimate as
\begin{equation}
\label{eq2:MLISIC}
\begin{split}
{{\hat a}_{2,k - 2L + 2}} &= {\rm{deci}}\left\{ {{y_{k - 2L + 2}}} \right.  \\
&- \left( {{G_{1,L}}{{\hat a}_{1,k - 3L + 3}} +  \ldots  + {G_{1,2}}{{\hat a}_{1,k - 2L + 1}}} \right) \\
&\left. { - \left( {{G_{1,2}}{{\hat a}_{1,k - 2L + 3}} +  \ldots  + {G_{1,L}}{{\hat a}_{1,k - L + 1}}} \right)} \right\}. \\
\end{split}
\end{equation}

Similarly, when the symbol ${y_k}$ is received, the $K'$-th ($ K' \ge 2$) estimation is given in (\ref{eq3:MLISIC}).

\begin{figure}[!h]
	\centering
	\includegraphics[width=8.8cm]{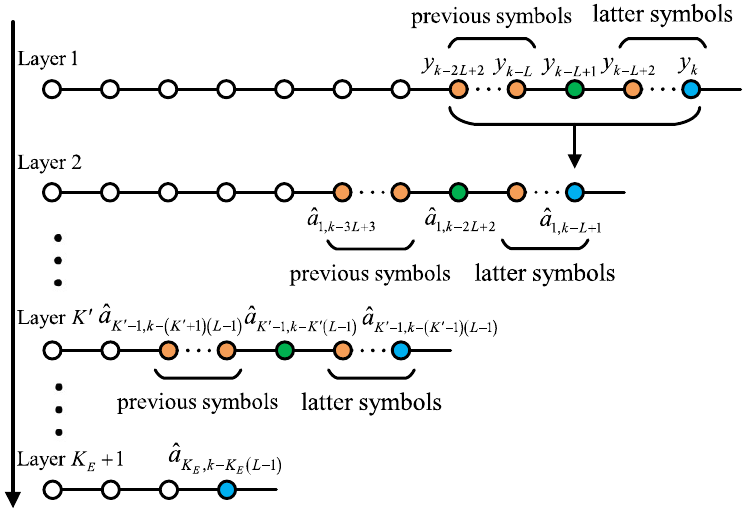}
	\caption{Complete estimation of the proposed MLISIC algorithm.}
	\label{MLISICAlgorithm}
\end{figure}

The complete estimation of the proposed MLISIC algorithm is shown in Fig. \ref{MLISICAlgorithm}, where ${K_E}$ indicates the number of iteration. Each circle in Fig. \ref{MLISICAlgorithm} represents a received or estimated symbol. The subscript of the green circle in a layer is consistent with that of the symbol re-estimated using symbols in the current layer. The orange and blue circles represent the symbols used for the estimation of the corresponding layer. In particular, the blue circle indicates the latest estimated symbol obtained by the current layer.

The proposed MLISIC algorithm is summarized in Algorithm \ref{alg:1}.
\begin{algorithm}[!h]
	\caption{Proposed MLISIC algorithm for FTN signaling.}
	\label{alg:1}
	\LinesNumbered 
	\KwIn{The received symbol ${y_k}$, the rolling factor $\alpha$ of SRRC shaping pulse, the time acceleration parameter $\tau $ for FTN signaling and the estimation parameters ${K_E}$ and $L$.}
	\KwOut{Symbols after the last estimation, i.e., symbols in the (${K_E} + 1$)-th layer.	}
	According to the given SRRC filter, calculate $\left[ {{G_{1,2}},{G_{1,3}}, \ldots {G_{1,L}}} \right]$\; 
	$i = 0$\; 
	\While{$i < {K_E}$}{
		\eIf{$i = 0$}{
			perform the first estimation with received symbols by using (\ref{eq1:MLISIC})\;
		}{
			symbols after the previous estimation are used to perform the $(i+1)$-th estimation as expressed in (\ref{eq3:MLISIC})\;
		}
		$i = i + 1$;
	}
\end{algorithm}

\subsection{Proposed IMLISIC}

The MLISIC algorithm takes advantage of symbols in the previous layer to estimate symbols in the current layer, which means the estimation of each layer is independent. However, symbols that have been estimated in the current layer are not fully used in the estimation of the current layer. In addition, the latest estimated symbol in each layer is not used for the estimation that is performed after receiving the next symbol. Therefore, the IMLISIC algorithm is proposed to further improve the estimation accuracy. In order to be consistent with the MLISIC algorithm, assume that the number of iteration is ${K_E}$ and the single-side symbol length used for the ISI elimination in the $K'$-th ($1 \le K' \le {K_E}$) iteration is ${L_{K'}} - 1$.

After ${y_k}$ is received, the first iteration of the proposed IMLISIC algorithm can be denoted as
\begin{equation}
\setcounter{equation}{15} 
\label{eq1:IMLISIC}
\begin{split}
{{\hat a}_{1,k - {L_1} + 1}} &= {\rm{deci}}\left\{ {{y_{k - {L_1} + 1}}} \right.  \\
&- \left( {{G_{1,{L_1}}}{{\hat a}_{1,k - 2{L_1} + 2}} +  \ldots  + {G_{1,2}}{{\hat a}_{1,k - {L_1}}}} \right) \\
&- \left. {\left( {{G_{1,2}}{y_{k - {L_1} + 2}} +  \ldots  + {G_{1,{L_1}}}{y_k}} \right)} \right\}.\; \\
\end{split}
\end{equation}
It should be noted that this formula eliminates the ISI by using both symbols after the first estimation and received symbols.

And then, ${\hat a_{1,k'}},k - {L_1} - {L_2} + 3 \le k' \le k - {L_1} + 1$ in the second layer and ${\hat a_{2,k'}},k - {L_{\rm{1}}} - 2{L_2} + 3 \le k' \le k - {L_1} - {L_2} + {\rm{1}}$ in the third layer can be used to estimate ${\hat a_{2,k - {L_1} - {L_2} + 2}}$ as
\begin{equation}
\label{eq2:IMLISIC}
\begin{split}
{\hat a}_{2,k} &_{- {L_1} - {L_2} + 2} = {\rm{deci}}\left\{ {{y_{k - {L_1} - {L_2} + 2}}} \right.  \\
&- \left( {{G_{1,{L_2}}}{{\hat a}_{2,k - {L_1} - 2{L_2} + 3}} +  \ldots  + {G_{1,2}}{{\hat a}_{2,k - {L_1} - {L_2} + 1}}} \right) \\
&\left. { - \left( {{G_{1,2}}{{\hat a}_{1,k - {L_1} - {L_2} + 3}} +  \ldots  + {G_{1,{L_2}}}{{\hat a}_{1,k - {L_1} + 1}}} \right)} \right\}. \\
\end{split}
\end{equation}

Correspondingly, the $K'$-th ($2 \le K' \le {K_E}$) estimation is denoted in (\ref{eq3:IMLISIC}).

Furthermore, the latest estimated symbol after the $K'$-th ($2 \le K' \le {K_E}$) estimation ${\hat a_{K',k - \sum\nolimits_{i = 1}^{K'} {\left( {{L_i} - 1} \right)} }}$ can be used to update the estimated symbols with the corresponding subscript $k - \sum\nolimits_{i = 1}^{K'} {\left( {{L_i} - 1} \right)} $ in the previous layers. Hence, ${L_{K'}} - 1$ ($1 \le K' \le {K_E}$) must be set elaborately. And ${L_{{K_E}}}$ should be set carefully to fully eliminate the ISI. The latest estimated symbol in the $K'$-th (${K_E} + 1 \ge K' \ge 3$) layer can contribute to the next estimation of the $K''$-th (${K_E} \ge K'' \ge 2,K' > K''$) layer if
\begin{equation}
\label{eq4:IMLISIC}
\begin{split}
{L_{K'' - 1}} - 1 &\ge k - \sum\nolimits_{i = 1}^{K'' - 1} {\left( {{L_i} - 1} \right)}  + 1  \\
& - \left( {k - \sum\nolimits_{i = 1}^{K' - 1} {\left( {{L_i} - 1} \right)} } \right) + 1 - 1 \\
& = \sum\nolimits_{i = K''}^{K' - 1} {\left( {{L_i} - 1} \right)}  + 1,
\setcounter{equation}{18} 
\end{split}
\end{equation}
where ${k - \sum\nolimits_{i = 1}^{K'' - 1} {\left( {{L_i} - 1} \right)} }$ and ${k - \sum\nolimits_{i = 1}^{K' - 1} {\left( {{L_i} - 1} \right)} }$ indicate subscripts of the latest estimated symbols in the ${K''}$-th and ${K'}$-th layers, respectively.

The optimal ${L_{K'}} - 1$ ($1 \le K' \le {K_E}$) should make sure that the latest estimated symbol in the $K'$-th (${K_E} + 1 \ge K' \ge 3$) layer can be used for the next estimation of all the previous layers. In this case, the relation between ${L_{K'}},K' \ne {K_E}$ and ${L_{{K_E}}}$ can be defined as
\begin{flalign}
\label{eq5:IMLISIC}
{L_{{K_E} - k}} \ge {2^{k - 1}}{L_{{K_E}}} + 1,1 \le k \le {K_E} - 1,
\end{flalign}
where the optimal length increases exponentially with the number of iteration ${K_E}$. Therefore, this formula is applicable to cases with small ${K_E}$ (generally, the satisfactory performance can be achieved when ${K_E}$ ranges from 2 to 6). In addition, it is not necessary to meet the optimal length, and the estimation accuracy can be sacrificed to simplify the complexity. One of these simplification methods can be represented as 
\begin{flalign}
\label{eq6:IMLISIC}
{L_{K' - 1}}{\rm{ = }}{L_{K'}} + 1,2 \le K' \le {K_E},
\end{flalign} 
with which the latest estimated symbol in the current layer only contributes to the next estimation of the previous layer.

\begin{figure}[!h]
	\centering
	\includegraphics[width=8.8cm]{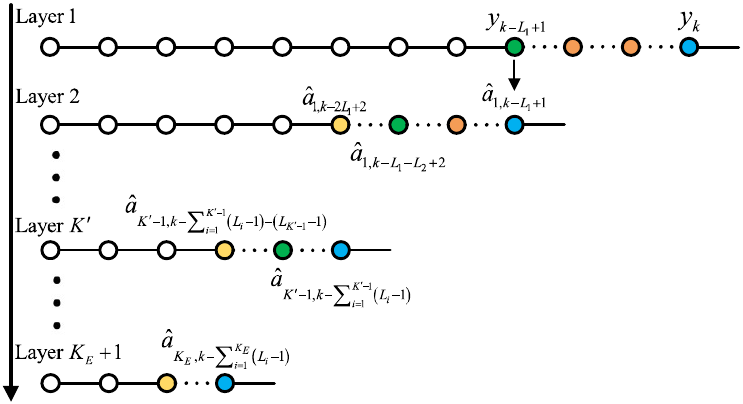}
	\caption{The complete estimation of the proposed IMLISIC algorithm.}
	\label{IMLISICAlgorithm}
\end{figure}

Similarly, the estimation diagram of the proposed IMLISIC algorithm is shown in Fig. \ref{IMLISICAlgorithm}. Circles in the first layer indicate the received symbols. Orange and blue circles represent the right symbols used for the estimation of the next layer. Yellow, green, and orange circles in the current layer indicate the left symbols used for the estimation of the current layer.

The proposed IMLISIC algorithm is summarized in Algorithm \ref{alg:2}.

\begin{algorithm}[!h]
	\caption{Proposed IMLISIC algorithm for FTN signaling.}
	\LinesNumbered 
	\label{alg:2}
	\KwIn{The received symbol ${y_k}$, the rolling factor $\alpha$ of SRRC shaping pulse, the time acceleration parameter $\tau $ for FTN signaling and the estimation parameters ${K_E}$ and $\left[ {{L_1},{L_2}, \ldots ,{L_{{K_E}}}} \right]$.}
	\KwOut{Symbols after the last estimation, i.e., symbols in the (${K_E} + 1$)-th layer.}
	According to the given SRRC filter, calculate $\left[ {{G_{1,2}},{G_{1,3}}, \ldots {G_{1,\hat L}}} \right],\hat L{\rm{ = arg}}\mathop {\max }\limits_{a \in \left[ {{L_1},{L_2}, \ldots {L_{{K_E}}}} \right]} a$\;
	$i = 0$\; 
	\While{$i < {K_E}$}{
		\eIf{$i = 0$}{
			perform the first estimation with the received symbols by using (15)\;
		}{
			perform the $(i+1)$-th estimation as expressed in (\ref{eq3:IMLISIC})\;
		}
		$i = i + 1$;
	}
\end{algorithm}

\subsection{Comparison of the Computational Complexity}

In order to eliminate the ISI on both sides of the desired symbol, the MLISIC algorithm needs to carry out $2\left( {L - 1} \right)$ additions/subtractions and $2\left( {L - 1} \right)$ multiplications in every estimation. Hence, given the estimation parameter ${K_E}$ and $L$, $2{K_E}\left( {L - 1} \right)$ additions/subtractions and $2{K_E}\left( {L - 1} \right)$ multiplications are needed. As for the IMLISIC algorithm, $2\left( {{L_{K'}} - 1} \right)$ additions/subtractions and $2\left( {{L_{K'}} - 1} \right)$ multiplications are required in the $K'$-th ($1 \le K' \le {K_E}$) estimation. The total computational complexity includes $2\sum\nolimits_{K' = 1}^{{K_E}} {\left( {{L_{K'}} - 1} \right)} $ additions/subtractions and $2\sum\nolimits_{K' = 1}^{{K_E}} {\left( {{L_{K'}} - 1} \right)} $ multiplications. Comparison with the algorithms in \cite{7886296} is shown in Table \uppercase\expandafter{\romannumeral1}.

\begin{table}[h]
\centering
\caption{The comparison of the computational complexity}
\renewcommand\arraystretch{1.3}
\begin{tabular}{l|cc}
  \hline
  Algorithm & \tabincell{c}{Number of \\ additions/subtractions} & \tabincell{c}{Number of \\ multiplications} \\
  \hline      
  SSSSE in \cite{7886296} & $L - 1$ & $L - 1$ \\
  \hline
  SSSgb$\emph{K}$SE in \cite{7886296} & \tabincell{c}{$\left( {K + 2} \right)\left( {L - 1} \right) + $\\${{K\left( {K + 1} \right)} \mathord{\left/ {\vphantom {{K\left( {K + 1} \right)} 2}} \right. \kern-\nulldelimiterspace} 2}$} & \tabincell{c}{$\left( {K + 2} \right)\left( {L - 1} \right) + $\\${{K\left( {K + 1} \right)} \mathord{\left/ {\vphantom {{K\left( {K + 1} \right)} 2}} \right. \kern-\nulldelimiterspace} 2}$} \\
  \hline  
  Proposed MLISIC & $2{K_E}\left( {L - 1} \right)$ & $2{K_E}\left( {L - 1} \right)$ \\
  \hline
  Proposed IMLISIC & $2\sum\nolimits_{K' = 1}^{{K_E}} {\left( {{L_{K'}} - 1} \right)} $ & $2\sum\nolimits_{K' = 1}^{{K_E}} {\left( {{L_{K'}} - 1} \right)} $ \\
\hline  
\end{tabular}
\end{table}

With the $\emph{U}$ random samples of the Gaussian randomization and the block length $J$, the algorithms presented in \cite{7990502} require the computational complexity of ${\rm O}\left( {{{\left( {4J + 1} \right)}^{3.5}} + {{\left( {4J + 1} \right)}^2}U} \right)$ and ${\rm O}\left( {{{\left( {2J + 1} \right)}^{3.5}} + {{\left( {2J + 1} \right)}^2}U} \right)$, respectively. These algorithms show satisfactory performance, whereas their computational complexities are too high for the practical implementation. 

\begin{figure}[!h]
	\centering
	\includegraphics[width=8.8cm]{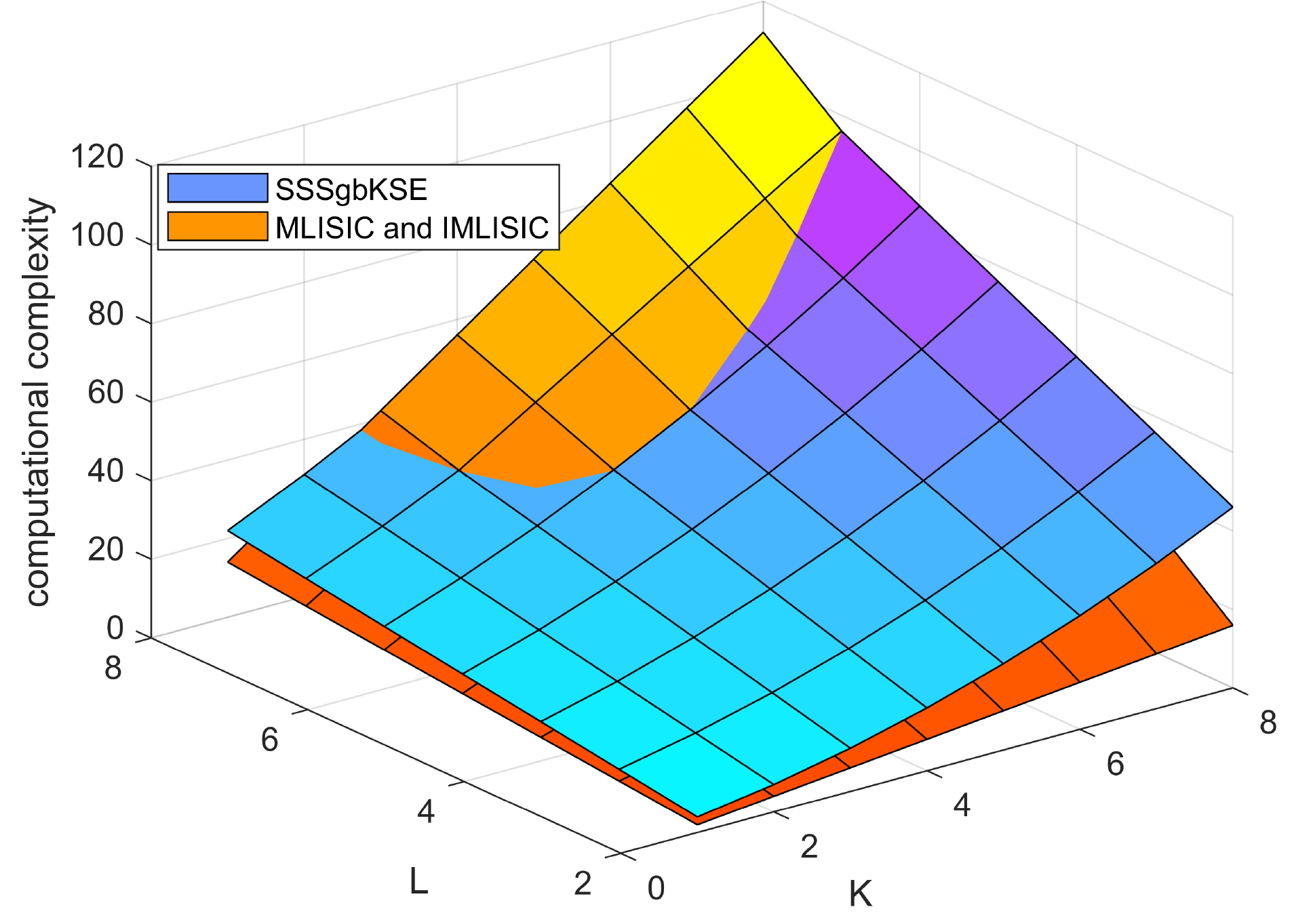}
	\caption{Multiplication computational complexity of SSSgb$\emph{K}$SE, MLISIC and IMLISIC with $K = {K_E},L = {L_{K'}},1 \le K' \le {K_E}$.}
	\label{complexityCompare}
\end{figure}

Besides, the multiplication numbers of the SSSgb$\emph{K}$SE algorithm and the proposed algorithms are shown in Fig. \ref{complexityCompare}, where $K = {K_E},L = {L_{K'}},1 \le K' \le {K_E}$. In this case, it can be seen that the multiplication numbers of the proposed algorithms are lower than that of the SSSgb$\emph{K}$SE algorithm when $\emph{K}$ or $\emph{L}$ is small (the most common cases), otherwise, our proposed algorithms require more complexity. Furthermore, the computational complexity of the IMLISIC algorithm is higher than that of the proposed MLISIC algorithm if ${L_{K'}} > L$. In particular, at least ${2^{{K_E}}}{L_{{K_E}}} - 2$ additions/subtractions and ${2^{{K_E}}}{L_{{K_E}}} - 2$ multiplications are necessary for the IMLISIC algorithm when ${L_{K'}}$ meets the optimal length, however, the computation complexity is still acceptable as long as ${K_E}$ and ${L_{{K_E}}}$ are not that large (common scenarios). The higher estimation accuracy can also be ensured even if ${L_{K'}}$ is decreased to reduce the complexity. In addition, we will show that our proposed MLISIC and IMLISIC algorithms can achieve better BER performance with lower computational complexity than the SSSgb$\emph{K}$SE algorithm in the next section. 

\section{Simulation Results}

Based on the FTN system model in Section \uppercase\expandafter{\romannumeral2}, the performance of the proposed MLISIC and IMLISIC algorithms is evaluated in this section. Quadrature phase shift keying
(QPSK), 8-phase shift keying (8-PSK) and 16/32/64/128/256-APSK are considered. Assuming that the order of SRRC filters is known and equals to 201, rolling factors $\alpha$ of 0.3, 0.4 and 0.5 are applied in the FTN system. Besides, we only consider the time acceleration factors of 0.8 and 0.9. Unless otherwise mentioned, the length of cyclic prefix for the FDE algorithm in \cite{6574905} is 10 and the block length equals to 1024, which means the spectrum efficiency is decreased by 0.98\%.

\begin{table}[h]
	\centering
	\caption{All the simulated cases in this paper}
	\renewcommand\arraystretch{1.3}
	\begin{tabular}{lcccc}
		\hline
		$\tau$ & $\eta$ & $\alpha$ & ISI type & \tabincell{c}{Simulated modulation} \\
		\hline     
		\multirow{2}{*}{0.8} & \multirow{2}{*}{25\%} & 0.3,0.4 & relatively severe ISI & QPSK, 8-PSK, 16-APSK \\
		\cline{3-5} 
		~ & ~ & 0.5 & moderate ISI & 16/32/64/128/256-APSK\\
		\hline
		0.9 & 11\% & 0.3 & mild ISI & \tabincell{c}{QPSK, 8-PSK, \\16/32/64/128/256-APSK} \\
		\hline  
	\end{tabular}
\end{table}

For the convenience of readers, all the simulated cases are summarized in Table \uppercase\expandafter{\romannumeral2}, where $\eta$ represents the improvement ratio of spectrum efficiency compared with the corresponding Nyquist system.

\subsection{Simulations in Mild ISI Cases}

\begin{figure}[!h]
	\centering
	\includegraphics[width=8.5cm]{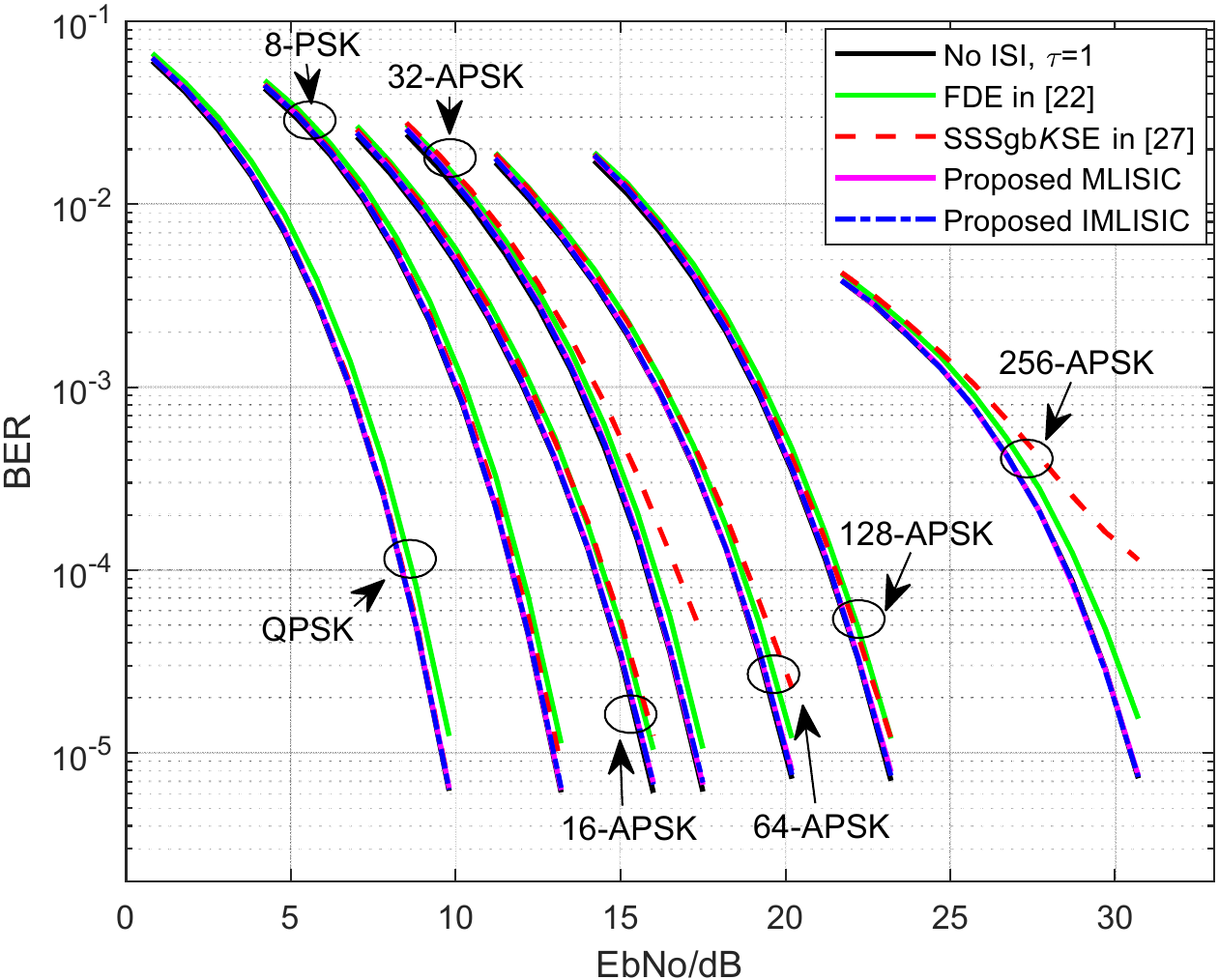}
	\caption{BER performance of the FTN system with $\tau  = 0.9,\alpha  = 0.3$.}
	\label{lowISI}
\end{figure}

BER performance in mild ISI cases with $\tau  = 0.9,\alpha  = 0.3$ is shown in Fig. \ref{lowISI}. Other simulation parameters can be seen in Table \uppercase\expandafter{\romannumeral3}, where $L$ is equivalent to $\left[ {{L_1},{L_2}, \ldots ,{L_{{K_E}}}} \right]$ for the IMLISIC algorithm.

\begin{table}[h]
	\centering
	\caption{Simulation parameters in mild ISI cases}
	\renewcommand\arraystretch{1.3}
	\begin{tabular}{cccccc}
		\hline
		Modulation & Algorithm & $L$ & $K\;or\;{K_E}$ \\
		\hline   
		\multirow{3}{*}{\tabincell{c}{QPSK \\8-PSK \\16-APSK}} & SSSgb$\emph{K}$SE & 6 & 3 \\
		\cline{2-4}   
		~ & MLISIC & 6 & 2 \\	
		\cline{2-4} 
		~ & IMLISIC & [7,6] & 2 \\	
		\hline
		\multirow{3}{*}{32-APSK} & SSSgb$\emph{K}$SE & 8 & 3 \\
		\cline{2-4}   
		~ & MLISIC & 8 & 2 \\	
		\cline{2-4} 
		~ & IMLISIC & [9,8] & 2 \\	
		\hline
		\multirow{3}{*}{64-APSK} & SSSgb$\emph{K}$SE & 8 & 4 \\
		\cline{2-4}   
		~ & MLISIC & 8 & 3 \\	
		\cline{2-4} 
		~ & IMLISIC & [8,8,8] & 3 \\	
		\hline
		\multirow{3}{*}{128-APSK} & SSSgb$\emph{K}$SE & 8 & 5 \\
		\cline{2-4}   
		~ & MLISIC & 8 & 4 \\	
		\cline{2-4} 
		~ & IMLISIC & [8,8,8,8] & 4 \\	
		\hline	
		\multirow{3}{*}{256-APSK} & SSSgb$\emph{K}$SE & 13 & 5 \\
		\cline{2-4}   
		~ & MLISIC & 13 & 4 \\	
		\cline{2-4} 
		~ & IMLISIC & [13,13,13,13] & 4 \\	
		\hline									
	\end{tabular}
\end{table}

Conclusions drawn from Fig. \ref{lowISI} are summarized as follows. Firstly, the proposed algorithms are superior to the SSSgb$\emph{K}$SE and FDE algorithms no matter which modulation is adopted. And it should be noted that according to Table \uppercase\expandafter{\romannumeral1} and Table \uppercase\expandafter{\romannumeral3}, the computational complexities of the simulated MLISIC and IMLISIC algorithms are lower than that of the SSSgb$\emph{K}$SE algorithm. In other words, our proposed MLISIC and IMLISIC algorithms can achieve better BER performance with lower computational complexity than the SSSgb$\emph{K}$SE and FDE algorithms in these cases. Secondly, it can be observed that the theoretical performance of all the simulated modulations (256-APSK included) can be approached with our proposed algorithms. The BER performance degradation is no more than 0.03 dB for 256-APSK when the theoretical BER is ${10^{ - 5}}$, and it could be lower when adopting lower-order modulations. By contrast, the FDE algorithm can not approximate the theoretical performance in the face of mild ISI cases, and the SSSgb$\emph{K}$SE algorithm can only achieve this performance when adopting low-order modulations, such as QPSK. Particularly, the performance degradation of the SSSgb$\emph{K}$SE algorithm is almost 2 dB when adopting 256-APSK. Compared with the SSSgb$\emph{K}$SE and FDE algorithms, therefore, there is no doubt that our proposed algorithms are more suitable for higher-order modulations in mild ISI cases. Furthermore, the performance difference between the proposed MLISIC and IMLISIC algorithms is tiny due to the mild ISI.

\subsection{Simulations in Moderate ISI Cases}

\begin{figure}[!h]
	\centering
	\includegraphics[width=8.5cm]{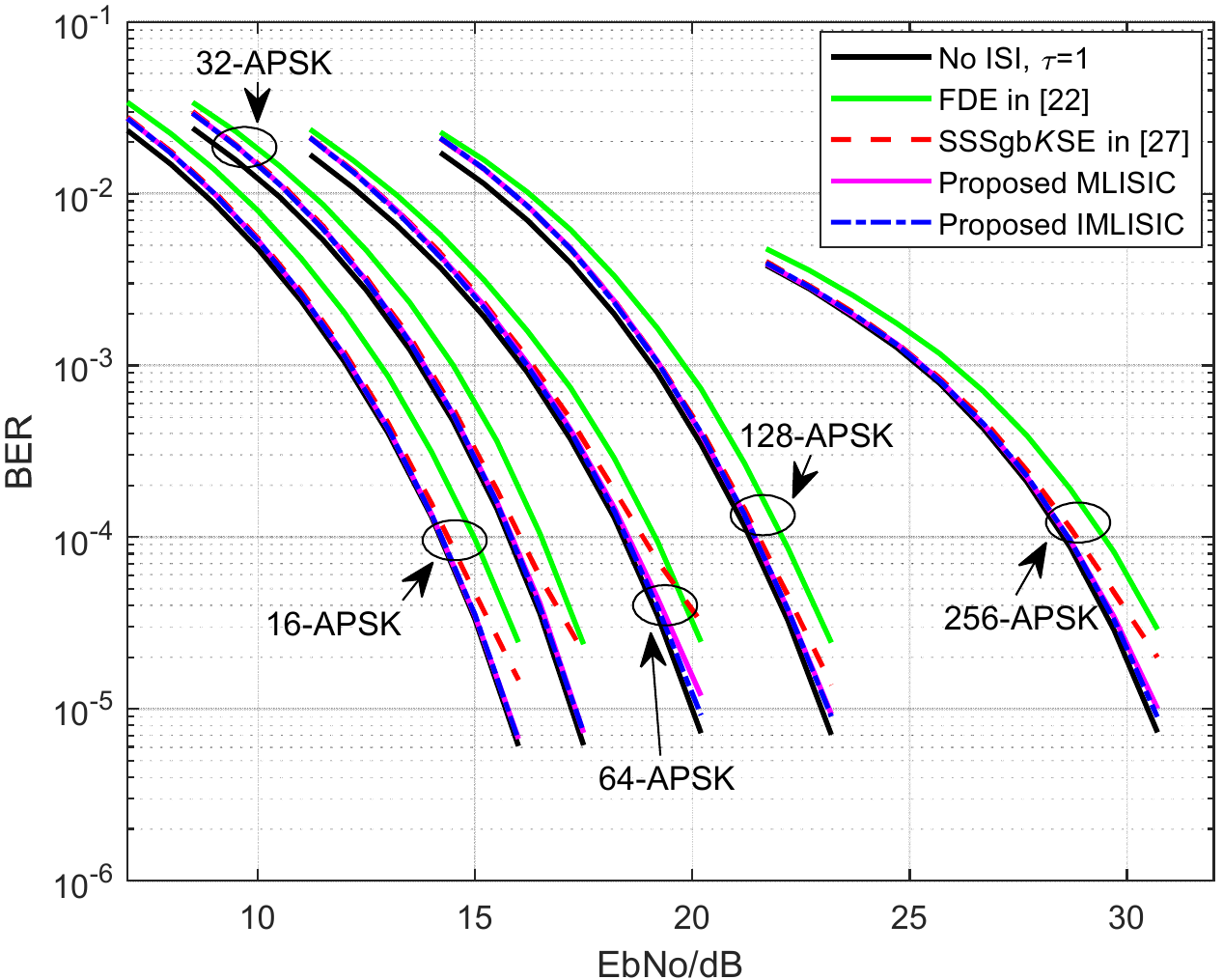}
	\caption{BER performance of the FTN system with $\tau  = 0.8,\alpha  = 0.5$.}
	\label{moderateISI}
\end{figure}

BER performance in moderate ISI cases with $\tau  = 0.8,\alpha  = 0.5$ is shown in Fig. \ref{moderateISI}. Other simulation parameters are listed in Table \uppercase\expandafter{\romannumeral4}.

\begin{table}[h]
	\centering
	\caption{Simulation parameters in moderate ISI cases}
	\renewcommand\arraystretch{1.3}
	\begin{tabular}{cccccc}
		\hline
		Modulation & Algorithm & $L$ & $K\;or\;{K_E}$ \\
		\hline   
		\multirow{3}{*}{16-APSK} & SSSgb$\emph{K}$SE & 8 & 4 \\
		\cline{2-4}   
		~ & MLISIC & 8 & 4 \\	
		\cline{2-4} 
		~ & IMLISIC & [13,7,6] & 3 \\	
		\hline
		\multirow{3}{*}{32/64-APSK} & SSSgb$\emph{K}$SE & 8 & 5 \\
		\cline{2-4}   
		~ & MLISIC & 8 & 5 \\	
		\cline{2-4} 
		~ & IMLISIC & [10,9,8,7,6] & 5 \\	
		\hline
		\multirow{3}{*}{128/256-APSK} & SSSgb$\emph{K}$SE & 8 & 6 \\
		\cline{2-4}   
		~ & MLISIC & 8 & 6 \\	
		\cline{2-4} 
		~ & IMLISIC & [13,12,11,10,9,8] & 6 \\	
		\hline									
	\end{tabular}
\end{table}

In contrast with the curves Fig. \ref{lowISI}, due to the severer ISI, the BER performance of most algorithms in Fig. \ref{moderateISI} becomes worse even though their complexities may be higher. In particular, BER performance of the SSSgb$\emph{K}$SE algorithm is much better when adopting 256-APSK because of the increase of $K$. As shown in Fig. \ref{moderateISI}, the FDE algorithm is quite disappointing in moderate ISI cases. Nevertheless, our proposed MLISIC and IMLISIC algorithms are still superior to the other two algorithms and achieve excellent performance in moderate cases. To be more specific, for the theoretical BER of ${10^{ - 5}}$, the performance degradation of the proposed MLISIC and IMLISIC algorithms is almost 0.06 dB when adopting 16/32-APSK. Using 64/128/256-APSK, the degradation of our proposed MLISIC and IMLISIC algorithms is about 0.13 dB and 0.19 dB, respectively. Certainly, the performance of our proposed algorithms can be further improved by increasing $L$ or ${K_E}$. In addition, there is almost no difference in BER performance among the SSSgb$\emph{K}$SE algorithm and our proposed algorithms when signal to noise ratio (SNR) is low.

\subsection{Simulations in relatively Severe ISI Cases}

\begin{figure}[!h]
	\centering
	\includegraphics[width=8.5cm]{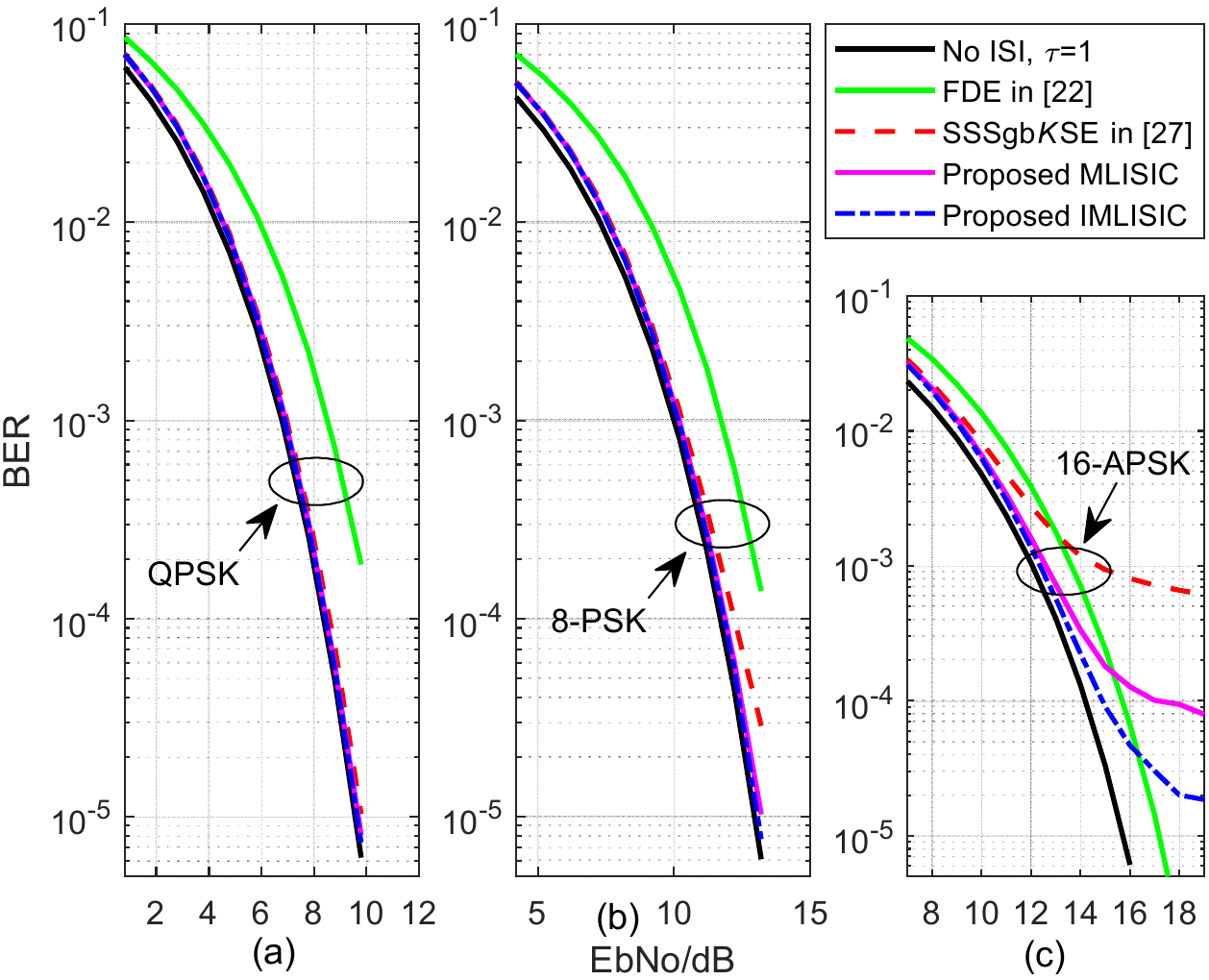}
	\caption{BER performance of the FTN system with $\tau  = 0.8,\alpha  = 0.4$.}
	\label{relaBadISI}
\end{figure}

\begin{figure}[!h]
	\centering
	\includegraphics[width=8.5cm]{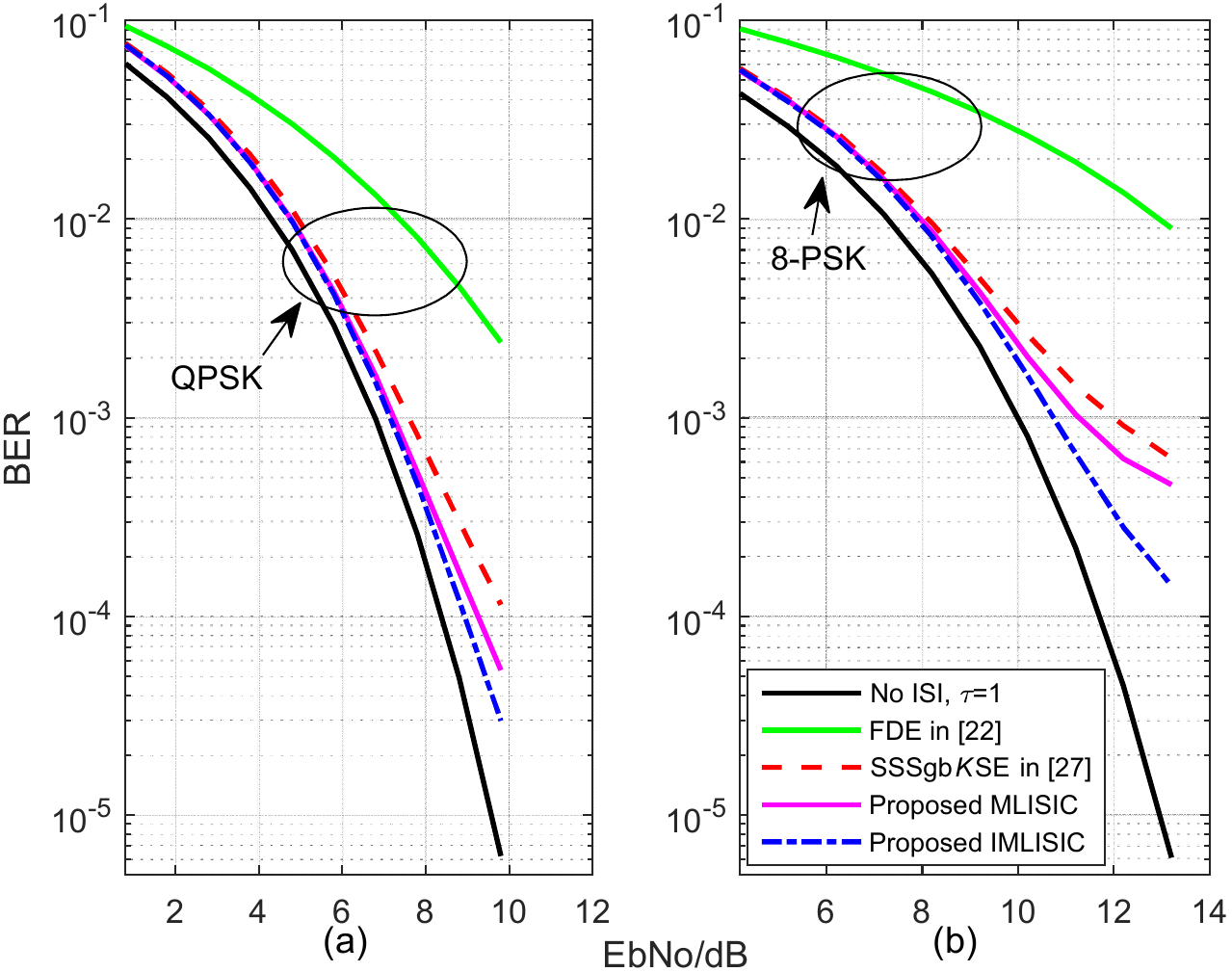}
	\caption{BER performance of the FTN system with $\tau  = 0.8,\alpha  = 0.3$.}
	\label{severeISI}
\end{figure}

BER curves in relatively severe ISI cases with $\tau  = 0.8,\alpha  = 0.4$ and $\alpha  = 0.3$ are shown in Fig. \ref{relaBadISI} and Fig. \ref{severeISI}, respectively. Other simulation parameters are listed in Table \uppercase\expandafter{\romannumeral5}. 

\begin{table}[h]
	\centering
	\caption{Simulation parameters in relatively severe ISI cases}
	\renewcommand\arraystretch{1.3}
	\begin{tabular}{cccccc}
		\hline
		Modulation & Algorithm & $L$ & $K\;or\;{K_E}$ \\
		\hline   
		\multirow{3}{*}{QPSK} & SSSgb$\emph{K}$SE & 6 & 3 \\
		\cline{2-4}   
		~ & MLISIC & 6 & 3 \\	
		\cline{2-4} 
		~ & IMLISIC & [8,7,6] & 3 \\	
		\hline
		\multirow{3}{*}{8-PSK} & SSSgb$\emph{K}$SE & 6 & 4 \\
		\cline{2-4}   
		~ & MLISIC & 6 & 4 \\	
		\cline{2-4} 
		~ & IMLISIC & [9,8,7,6] & 4 \\	
		\hline
		\multirow{3}{*}{16-APSK} & SSSgb$\emph{K}$SE & 8 & 4 \\
		\cline{2-4}   
		~ & MLISIC & 8 & 4 \\	
		\cline{2-4} 
		~ & IMLISIC & [25,13,7,6] & 4 \\	
		\hline									
	\end{tabular}
\end{table}

Through the observation of Fig. 8(a) and Fig. 8(b), BER performance of our proposed algorithms is still acceptable and better than that of the SSSgb$\emph{K}$SE and FDE algorithms. As shown in Fig. 8(a), the degradation of our proposed MLISIC and IMLISIC algorithms is almost 0.07 dB and 0.12 dB for QPSK, respectively. As for 8-PSK, the corresponding degradation is approximately 0.1 dB and 0.22 dB. When adopting the SSSgb$\emph{K}$SE algorithm and our proposed algorithms as shown in Fig. 8(c), error floor appears, which means BER can not decline as SNR increases. Even in this case, the error floor can be reduced with our proposed algorithms. There is no error floor with the FDE algorithm, which shows the advantage of this algorithm to some extent.


Similarly, it can be seen from Fig. \ref{severeISI} that our proposed algorithms remain the superiority to the other two algorithms. Furthermore, BER performance of the IMLISIC algorithm is better than that of the MLISIC algorithm. However, it can be concluded from the results that the performance of these algorithms still has a lot of room for improvement, especially for high-order modulations.

\subsection{Analysis of the Proposed MLISIC and IMLISIC Algorithms}

\begin{figure}[!h]
	\centering
	\includegraphics[width=8.5cm]{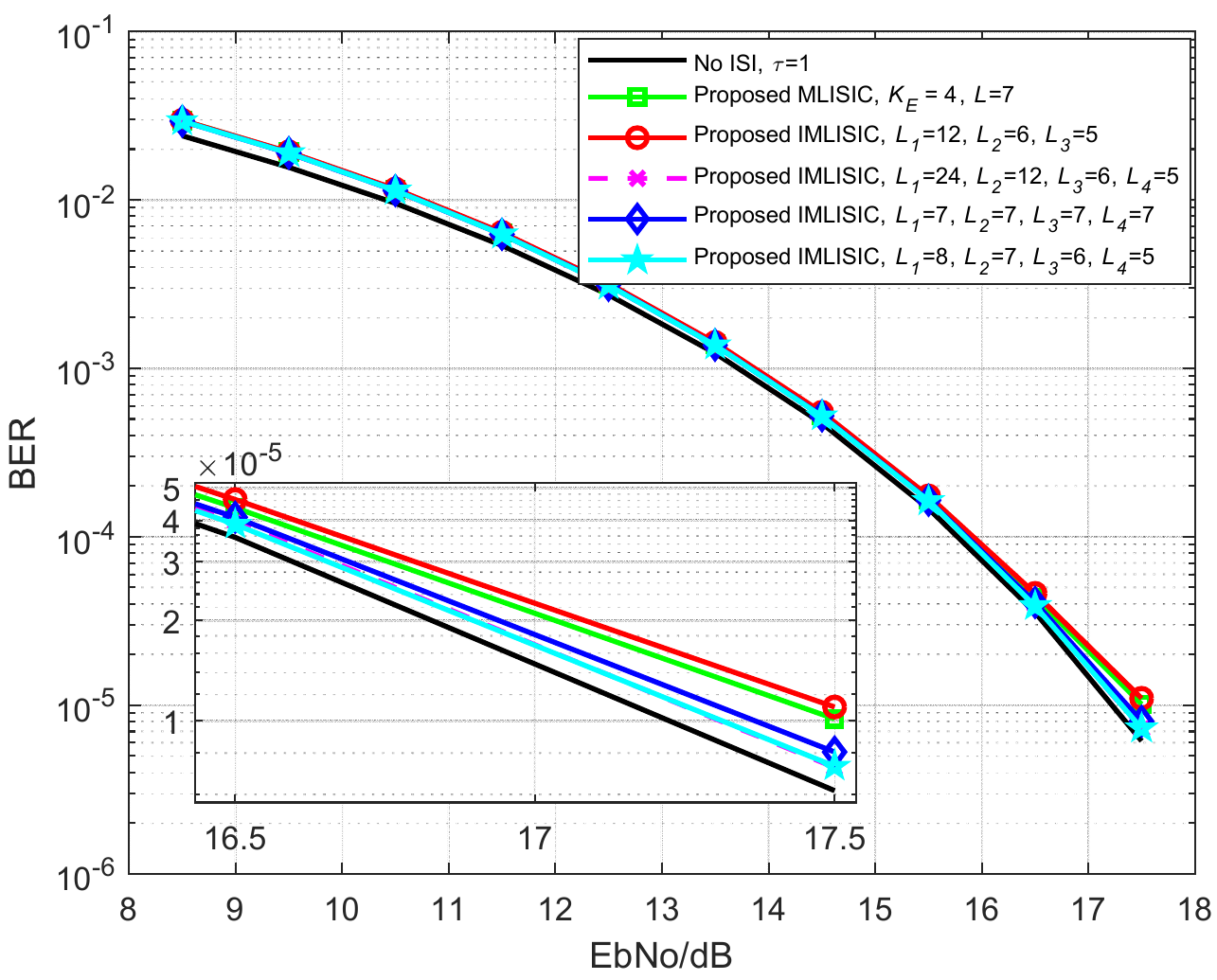}
	\caption{BER performance comparison of the 32-APSK FTN system between the proposed MLISIC and IMLISIC algorithms with $\tau {\rm{ = }}0.8,\alpha  = 0.5$.}
	\label{a32pskAnalyse}
\end{figure}

To further illustrate the proposed MLISIC and IMLISIC algorithms, their BER performance for the 32-APSK FTN system with $\tau {\rm{ = }}0.8,\alpha  = 0.5$ is shown in Fig. \ref{a32pskAnalyse}. It's worth mentioning that although only one situation is demonstrated here, the conclusions are universal. And the parameter $\tau {\rm{ = }}0.8,\alpha  = 0.5$ is chosen just because the phenomenon is more obvious. Comparing the green and blue curves, the superiority of the IMLISIC algorithm to the MLISIC algorithm can be verified. Besides, it can be concluded that increasing the number of iteration ${K_E}$ improves BER performance by comparison between the red and magenta curves. The magenta, blue and cyan curves indicate that in terms of the IMLISIC algorithm, BER performance degradation caused by the simplified ${L_{K'}},1 \le K' \le {K_E}$ is negligible if ISI is not quite severe.

\section{Conclusion}

As a non-orthogonal transmission scheme, FTN signaling is a quite promising technology to alleviate the pressure brought by limited spectrum resources. Inspired by existing successive symbol-by-symbol sequence estimators, we propose a low complexity and high accuracy sequence estimator, i.e., MLISIC, which requires $2{K_E}\left( {L - 1} \right)$ additions/subtractions and $2{K_E}\left( {L - 1} \right)$ multiplications for one single estimation. Besides, based on the MLISIC algorithm, the IMLISIC algorithm, which utilizes more accurate symbols and elaborate length of successive interference cancellation, is proposed to further improve the estimation accuracy. Although its computational complexity is higher than that of the MLISIC algorithm in most cases, it is superior to the MLISIC algorithm even when their complexities are equal. Compared with most block-based algorithms, the computational complexities of the proposed MLISIC and IMLISIC algorithms are much lower. In addition, the performance of the proposed algorithms is superior to that of the SSSgb$\emph{K}$SE algorithm under all the simulation cases considered in this paper. Furthermore, for all the modulation types adopted in modern satellite communication standards such as DVB-S2 and DVB-S2X, our proposed algorithms can approximate the theoretical performance in mild ISI cases, which is far beyond the performance of the FDE algorithm. Even under moderate ISI circumstances, the proposed algorithms perform satisfactorily.

\bibliographystyle{IEEEtran}
\bibliography{Reference}


%

%
%
%
%
%

\ifCLASSOPTIONcaptionsoff
  \newpage
\fi

\end{document}